# Performance Evaluation of Orthogonal Frequency Division Multiplexing (OFDM) based Wireless Communication System with implementation of Least Mean Square Equalization technique

*Farhana Enam*
Assistant Professor
Dept. of Information & Communication Engineering
University of Rajshahi, Rajshahi,
Bangladesh

*Md. Ashraful Islam*
Lecturer
Dept. of Information & Communication Engineering
University of Rajshahi, Rajshahi, Bangladesh
e-mail: ras5615@gmail.com

*Md. Arif Rabbani*
Dept. of Information & Communication Engineering
University of Rajshahi, Rajshahi,
Bangladesh

*Sohag Sarker*
Dept. of Information & Communication Engineering
University of Rajshahi, Rajshahi,
Bangladesh

*Abstract—* **Orthogonal Frequency Division Multiplexing (OFDM) has recently been applied in wireless communication systems due to its high data rate transmission capability with high bandwidth efficiency and its robustness to multi-path delay. Fading is the one of the major aspect which is considered in the receiver. To cancel the effect of fading, channel estimation and equalization procedure must be done at the receiver before data demodulation. This paper mainly deals with pilot based channel estimation techniques for OFDM communication over frequency selective fading channels. This paper proposes a specific approach to channel equalization for Orthogonal Frequency Division Multiplex (OFDM) systems. Inserting an equalizer realized as an adaptive system before the FFT processing, the influence of variable delay and multi path could be mitigated in order to remove or reduce considerably the guard interval and to gain some spectral efficiency. The adaptive algorithm is based on adaptive filtering with averaging (AFA) for parameter update. Based on the development of a model of the OFDM system, through extensive computer simulations, we investigate the performance of the channel equalized system. The results show much higher convergence and adaptation rate compared to one of the most frequently used algorithms - Least Mean Squares (LMS)**

## I. INTRODUCTION

Multimedia wireless services require high data-rate transmission over mobile radio channels. Orthogonal Frequency Division Multiplexing (OFDM) is widely considered as a promising choice for future wireless communications systems due to its high-data-rate transmission capability with high bandwidth efficiency. In OFDM, the entire channel is divided into many narrow subchannels, converting a frequency-selective channel into a collection of frequency-flat channels[1].Moreover, intersymbol interference (ISI) is avoided by the use of cyclic prefix (CP), which is achieved by extending an OFDM symbol with some portion of its head or tail [2]. In fact, OFDM has been adopted in digital audio broadcasting (DAB), digital video broadcasting (DVB), digital subscriber line (DSL), and wireless local area network (WLAN) standards such as the IEEE 802.11a/b/g/n [3–6]. It has also been adopted for wireless broadband access standards such as the IEEE 802.16e [7, 8, 9], and as the core technique for the fourth-generation (4G) wireless mobile Communications [10].To eliminate the need for channel estimation and tracking, Quadrature phase-shift keying (QPSK) can be used in OFDM systems. However, this result in a 3 dB loss in signal-to-noise ratio (SNR) compared with coherent demodulation such as phase-shift keying (PSK) [11]. The performance of OFDM systems can be improved by allowing for coherent demodulation when an accurate channel estimation technique is used. Channel estimation techniques for OFDM systems can be grouped into two categories: blind and non-blind. These blind channel estimation techniques may be a desirable approach as they do not require training or pilot signals to increase the system bandwidth and the channel throughput; they require, however, a large amount of data in order to make a reliable stochastic estimation. Therefore they suffer from high computational complexity and severe performance degradation in fast fading channel [12, 13, 14] . On the other hand, the non-blind channel estimation can be performed by either inserting pilot tones into all of the subcarriers of OFDM symbols with a specific period or inserting pilot tones into some of the subcarriers for each OFDM symbol [18, 19].In case of the non-blind channel estimation, the pilot tones are multiplexed with the data within an OFDM symbol and it is referred to as comb-type pilot arrangement. The comb-type channel estimation is performed to satisfy the need for the channel equalization or tracking in fast fading scenario, where the channel changes even in one OFDM period [15]. The main idea in comb-type channel estimation is to first estimate the channel conditions at the pilot subcarriers and then estimates the channel at the data subcarriers by means of interpolation. The estimation of the channel at the pilot subcarriers can be based on Least Mean-Square (LMS)[17].

The paper organizes as follows: section 2 describes the LMS(Least Mean Square) algorithm which is used in this research. The proposed model is described in section 3 as system description. In section 4, the simulation results of the proposed Least Mean Square equalization technique system are presented.

## II.  LMS(LEAST MEAN SQUARE) ALGORITHM

The Least Mean Square (LMS) algorithm is a gradient-based method of steepest decent [20].  LMS algorithm uses the estimates of the gradient vector from the available data. LMS incorporates an iterative procedure that makes successive corrections to the weight vector in the direction of the negative of the gradient vector which eventually leads to the minimum mean square error [21-24]. Compared to other algorithms LMS algorithm is relatively simple; it does not require correlation function calculation nor does it require matrix inversions.

Consider a Uniform Linear Array (ULA) with N isotropic elements, which forms the integral part of the adaptive beamforming system as shown in the figure below.

The output of the antenna array x(t) is given by,

$$X(t) = s(t)a(\theta_0) + \Sigma u(t)a(\theta_i) + n(t)$$

where, s(t) denotes the desired signal arriving at angle $\theta_0$ and $u_i(t)$ denotes interfering signals arriving at angle of incidences $\theta_i$ respectively, $a(\theta_0)$ and $a(\theta_0)$ represents the steering vectors for the desired signal and interfering signals respectively.

Therefore it is required to construct the desired signal from the received signal amid the interfering signal and additional noise n(t).

From the method of steepest descent, the weight vector equation is given by [16],

$$W(n+1) = w(n) + \tfrac{1}{2}\,\mu[-\Delta(E\{e^2(n)\})]$$

Where μ is the step-size parameter and controls the convergence chachteristics of the LMS algorithm; $e^2(n)$ is the mean square error between the beamformer output y(n) and the reference signal which is given by,

$$e^2(n) = [d^h(n) - w^h x(n)]^2$$

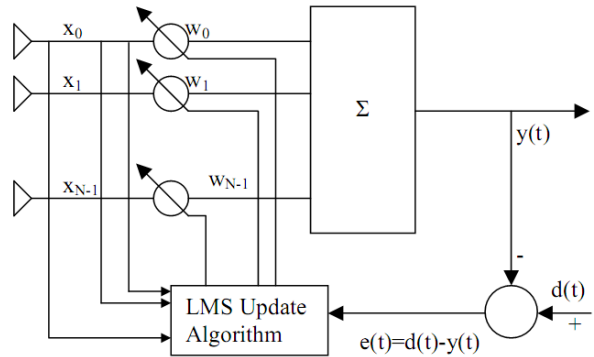

Figure 1: LMS adaptive beamforming network

The gradient vector in the above weight update equation can be computed as

$$\nabla_w\,(E\{e^2(n)\}) = -\,2r + 2Rw(n)$$

In the method of steepest descent the biggest problem is the computation involved in finding the values r and R matrices in real time. The LMS algorithm on the other hand simplifies this by using the instantaneous values of covariance matrices r and R instead of their actual values i.e.

$$R(n) = x(n)x^h(n)$$

$$r(n) = d^*(n)x(n)$$

Therefore the weight update can be given by the following equation,

$$w(n+1) = w(n) + \mu x(n)[d^*(n) - x^h(n)w(n)\,] = w(n) + \mu x(n)e^*(n)$$

The LMS algorithm is initiated with an arbitrary value w(0) for the weight vector at n=0. The successive corrections of the weight vector eventually leads to the minimum value of the mean squared error [25, 26]. Therefore the LMS algorithm can be summarized in following equations:

$$\text{Output, } y(n) = w^h x(n)$$
$$\text{Error, } e(n) = d^*(n) - y(n)$$

$$\text{Weight, } w(n+1) = w(n) + \mu x(n)e^*(n)$$

29

### III.  SIMULATION MODEL

In this section, the Wireless Communication system simulation with Least Mean Square (LMS) equalization technique   model to be implemented has been discussed thoroughly and all related assumptions have been stated clearly and justified. The implemented model needs to be realistic as possible in order to get reliable results. It is ought to be mentioned here that the real communication systems are very much complicated  and due to non availability of the algorithms to simulate the performance evaluation of their various sections, generally, simulations are made  on the basis of some   assumptions to simplify the   communication system(s)  concerned.

Figure-2 shows a simulation model for the Wireless Communication   system simulation with Least Mean Square (LMS) equalization technique. It consists of various sections. A brief description of the simulated model is given below:

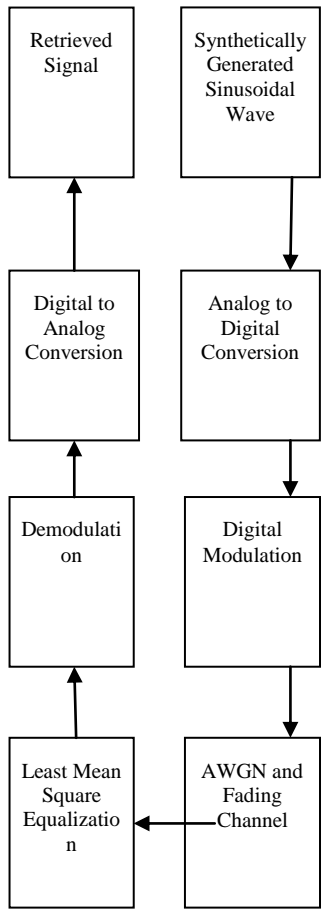

Figure-2: A block diagram of Wireless Communication system simulation with Least Mean Square (LMS) equalization technique.

The block diagram of the simulated system model is shown in Figure -2. The synthetically generated sinusoidal wave is first converted to digital bit stream. The digital signal is then fed to the input of the modulator. Then the data are modulated according to QPSK modulation scheme. The effect of AWGN and fading Channel are then introduced into the modulated wave. A Least Mean Square (LMS) equalization technique is used to remove the effect of AWGN and Fading channel. The output of the equalizer is then fed to the input of demodulator where the demodulation is done. Finally, the demodulated signal is converted to analog signal as the retrieved sinusoidal signal.

Table 1: The parameters of simulation model.:

| Parameters | values |
|---|---|
| Number Of Bits | 44000 |
| Number Of Subscribers | 200 |
| FFT Size | 256 |
| CP | 1/4 |
| Coding | Convolutional Coding(CC), Reed-Solomon(RS) Coding |
| Constraint length | 7 |
| K-factor | 3 |
| Maximum Doppler shift | 100/40Hz |
| Modulation | 16-QAM, 64-QAM, 256-QAM, QPSK, 16-PSK, 64-PSK, 256-PSK |
| Frequency used for synthetic data | 1 KHz |
| Sampling Rate | 4 KHz |
| SNR | 0-50 dB |
| Wireless channel | AWGN and Fading Channel |
| Channel Coefficients | [.986; .845; .237; .123+.31i] |

### IV.  SIMULATION RESULT

This section of the chapter presents and discusses all of the results obtained by the computer simulation program written in Matlab7.5, following the analytical approach of a wireless communication system considering AWGN and Fading channel. A test case is considered with the synthetically generated data. The results are represented in terms of bit energy to noise power spectral density ratio (Eb/No) and bit error rate (BER) for practical values of system parameters.

By varying SNR, the plot of  Eb/No vs BER  was  drawn with  the  help  of  "semilogy" function. The Bit Error Rate (BER) plot obtained in the performance analysis showed that model works well on Signal to Noise Ratio (SNR) less than 50 dB. Simulation results in figure-3 and figure-4 shows the performance of the system over AWGN and fading channels using QPSK, 16-PSK, 64-PSK, 256-PSK, 16-QAM, 64-QAM and 256-QAM modulation schemes respectively.

From figure-2, it is observed that the BER performance of the system with implementation of Least Mean Square algorithm (LMS) in QPSK outperforms as compared to other digital modulations. The system shows worst performance in 256 PSK. For a typical SNR value of 10dB, the system performance is improved by 8.26 dB. It is also noticeable that the system performance degrades with increase of order of modulation.

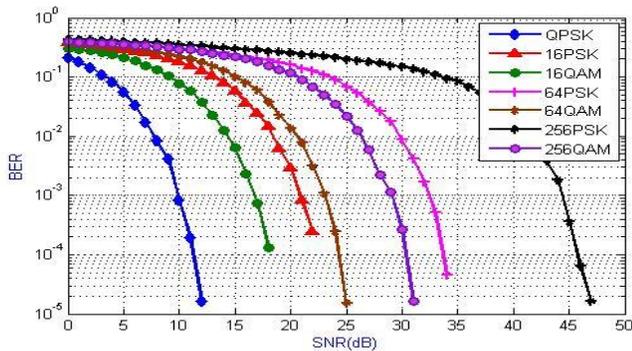

**Figure-2:** BER performance of a Wireless Communication System fewer than seven types of digital modulations over AWGN channel**.**

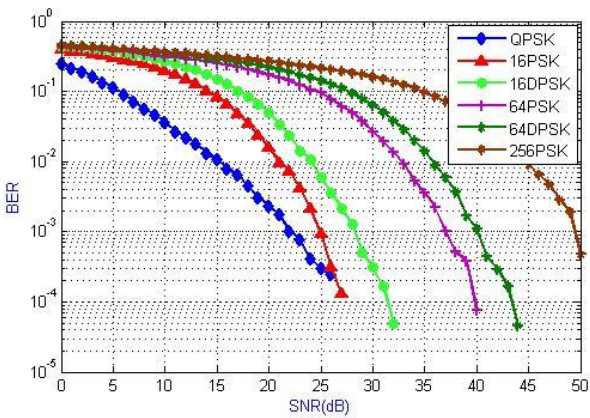

**Figure-3:** BER performance of a Wireless Communication System under different modulation schemes under fading channel

Figure-3 shows the BER performance of a Wireless Communication System under different modulation schemes under fading channel. From figure-3, it is also observed that the BER performance of the system with implementation of Least Mean Square algorithm (LMS) in QPSK is better as compared to other digital modulations. The system shows worst performance in 256 PSK. For a typical SNR value of 10dB, the system performance is improved by 4.511dB. It is also noticeable that the system performance degrades with increase of order of modulation.

## V. CONCLUTION

In this research work, it has been studied the performance of an OFDM based wireless communication system with implementation of Least Mean square equalization technique and different digital modulation schemes. A range of system performance results highlights the impact of digital modulations in AWGN and fading channels. From the present study it is found that the system performance is improved 7.36dB for QPSK modulation at SNR 9dB, 7.113dB for 64PSK modulation at SNR 31dB, 12.04dB for 64QAM modulation at SNR 22dB, 3.375dB for 16QAM modulation at SNR 15dB and 3.988dB for 256PSK modulation than uncoded situation over AWGN channel.

In the case of fading channel the system performance is improved 4.511dB for QPSK modulation at SNR 10dB, 7.203dB for 16PSK modulation at SNR 25dB, 7.964dB for 64PSK modulation at SNR 37dB, 6.0588dB for 16PSK modulation at SNR 22dB and 11.18dB for 256PSK modulation at SNR 47dB than uncoded situation.

In the present study, it has been observed that the OFDM, an elegant and effective multi carrier technique sed FEC encoded wireless communication system can overcome multipath distortion. In Bangladesh, WiMAX technology is going to be implemented and its physical layer is based on OFDM. The present work can be extended in MIMO-OFDM technology to ensure high data rate transmission.